\documentclass[conference]{IEEEtran}
\IEEEoverridecommandlockouts
\usepackage{cite}
\usepackage{amsmath,amssymb,amsfonts}
\usepackage{algorithmic}
\usepackage{graphicx}
\usepackage{algorithm}
\usepackage{textcomp}
\usepackage{subfigure}
\usepackage{xcolor}
\def\BibTeX{{\rm B\kern-.05em{\sc i\kern-.025em b}\kern-.08em
    T\kern-.1667em\lower.7ex\hbox{E}\kern-.125emX}}

\newtheorem{example}{Example}
\newtheorem{theorem}{\underline{Theorem}}

\newcommand{\mv}[1]{\mbox{\boldmath{$ #1 $}}}
\begin{document}

\title{Exploiting NOMA for Multi-Beam UAV Communication in Cellular Uplink}
\author{\IEEEauthorblockN{Liang Liu, Shuowen Zhang, and Rui Zhang}
\IEEEauthorblockA{ECE Department, National University of Singapore. Email: \{eleliu,elezhsh,elezhang\}@nus.edu.sg}}
\maketitle

\begin{abstract}
Unmanned aerial vehicles (UAVs) are expected to be an important new class of users in the fifth generation (5G) and beyond 5G cellular networks. In particular, there are emerging UAV applications such as aerial photograph and data relaying that require high-speed communications between the UAVs and the ground base stations (GBSs). Due to the high UAV altitude, the strong line-of-sight (LoS) links generally dominate the channels between the UAVs and GBSs, which brings both opportunities and challenges in the design of future wireless networks supporting both terrestrial and aerial users. Although each UAV can associate with more GBSs for communication as compared to terrestrial users thanks to the LoS-dominant channels, it also causes/suffers more severe interference to/from the terrestrial communications in the uplink/downlink. This paper studies the uplink communication from a multi-antenna UAV to a set of GBSs within its signal coverage by considering a practical yet challenging scenario when the number of antennas at the UAV is smaller than that of co-channel GBSs. To achieve high-rate transmission yet avoid interfering with any of the existing terrestrial communications at the co-channel GBSs, we propose a novel multi-beam transmission strategy by exploiting the non-orthogonal multiple access (NOMA) technique. Specifically, the UAV sends each data stream to a selected subset of the GBSs, which can decode the UAV's signals and then cancel them before decoding the messages of their served terrestrial users, and in the meanwhile nulls its interference at the other GBSs via zero-forcing (ZF) beamforming. To draw essential insight, we first characterize in closed-form the degrees-of-freedom (DoF) achievable for the UAV's sum-rate maximization under the proposed strategy. Then, we propose an efficient algorithm to jointly optimize the number of UAV data streams, the data stream-GBS association, and the transmit beamforming to maximize the UAV's transmit rate subject to the interference avoidance constraints for protecting the terrestrial users. Numerical examples are provided to verify the effectiveness of the proposed NOMA-based multi-beam transmission strategy.
\end{abstract}

\section{Introduction}\label{sec:Introduction}
Driven by the increasingly prosperous market of unmanned aerial vehicles (UAVs) for assorted applications such as traffic control, cargo delivery, surveillance, aerial inspection, etc., cellular-enabled UAV communication has drawn significant interests recently for realizing high-performance and beyond visual line-of-sight (BVLoS) UAV-ground communications \cite{Zhang18,LTE,CellularConnected}. As compared to conventional terrestrial communications, one distinct feature of UAV-ground communications is the dominant line-of-sight (LoS) component in the channels between the UAV and the ground base stations (GBSs) due to the high altitude of UAVs \cite{Qualcomm}, which brings both opportunities and challenges in the design of cellular networks. On one hand, the enhanced \emph{macro-diversity} in association with a large number of GBSs with strong LoS channels with the UAV can be exploited to improve the UAV communication performance \cite{Zhang18}, especially when the UAV is equipped with multiple antennas. However, on the other hand, severe \emph{co-channel interference} also occurs between the UAV and the GBSs that are serving the terrestrial users over the same frequency band, which may drastically degrade the communication performances of the terrestrial users in the uplink and the UAVs in the downlink \cite{Azari17,Nguyen18}, thus calling for more effective air-ground interference mitigation techniques \cite{Liang18,Mei18}.

To exploit the macro-diversity gain, this paper considers the scenario where a multi-antenna UAV sends multiple data streams to multiple single-antenna (or fixed-beam multi-antenna) GBSs over a given time-frequency resource block (RB) assigned, termed as multi-beam uplink transmission. In practice, there are a large number of GBSs in the UAV's signal coverage region due to the dominant LoS channels, which can be divided into two groups \cite{Liang18}: \emph{occupied} GBSs each receiving the signal from an associated terrestrial user in the same RB as the UAV, and \emph{available} GBSs which do not serve any terrestrial users in the uplink over this RB. In order to protect the existing terrestrial communications from the UAV's strong interference, we consider a stringent requirement on the UAV's multi-beam uplink transmission such that its uplink interference to all the occupied GBSs needs to be completely avoided. The conventional way to achieve this goal is to send the UAV data streams to the available GBSs only, and at the same time null the interference to all the occupied GBSs via zero-forcing (ZF) beamforming. However, since the number of occupied GBSs may be practically larger than the number of antennas at the UAV due to its space and hardware limitations, the ZF beamforming design can be infeasible.

Motivated by the rapid advance of non-orthogonal multiple access (NOMA) technologies (see e.g., \cite{NOMA,Ding2017,Ding2018} and the references therein), in this paper we aim to tackle the above challenge by optimizing the multi-beam transmission of the UAV such that a properly selected subset of the occupied GBSs can first decode the signals from the UAV and then cancel them for decoding the messages from their respectively served terrestrial users. Since the ZF beamforming vector for sending each data stream merely needs to be in the null space of the channels of the occupied GBSs that cannot decode this data stream, the ZF beamforming solution becomes more feasible as compared to the conventional ZF approach without applying NOMA or interference cancellation at the GBSs.

It is interesting to note that under our proposed scheme, available GBSs cannot be utilized for improving the transmit rate of the UAV when the number of occupied GBSs is larger than that of the antennas at the UAV (e.g., in a high terrestrial traffic load scenario). This is because each data stream has to be sent to some occupied GBSs for interference cancellation, thus it is not beneficial to send these data streams to any available GBSs for decoding. To address this challenging case, in this paper we focus on the system setting with occupied GBSs only. First, we characterize in closed-form the maximum degrees-of-freedom (DoF) of the UAV's achievable rate under the proposed multi-beam transmission scheme with NOMA. The obtained result reveals the maximum number of data streams that can be transmitted without degrading the performance of terrestrial communications after interference cancellation at infinite signal-to-noise ratio (SNR). Next, based on the DoF result, we devise an efficient algorithm in the finite SNR regime to jointly optimize the data stream-GBS associations as well as the transmit beamforming vectors and the rates of different data streams to maximize the achievable sum-rate of the UAV subject to the stringent interference avoidance constraints for protecting the terrestrial users.

It is worth pointing out that multi-beam UAV uplink transmission has been studied in our prior work \cite{Liang18}, where a different interference cancellation approach from NOMA, called cooperative interference cancellation, was proposed. Specifically, we assume in \cite{Liang18} that backhaul links between available GBSs and occupied GBSs exist such that the available GBSs can send their decoded messages from the UAV to the occupied GBSs via the backhaul links for interference cancellation. Theoretically, the cooperative interference cancellation strategy outperforms the NOMA-based strategy proposed in this paper with local interference cancellation at occupied GBSs only. However, this scheme critically relies on the message forwarding via backhaul links between GBSs, which may not be implementable in existing systems due to the overhead and decoding delay considerations. Therefore, we are motivated to consider NOMA requiring local interference cancellation only in this paper for ease of practical implementation.

The rest of this paper is organized as follows. Section \ref{sec:System Model} describes the system model for our considered multi-beam UAV communication in cellular uplink.
Section \ref{sec:NOMA-based Strategy} introduces the proposed NOMA-based interference avoidance strategy to protect the occupied GBSs.
Section \ref{sec:Problem Formulation} formulates the sum-rate maximization problem under the interference avoidance constraints.
Section \ref{sec:DoF Analysis} characterizes the maximum DoF of the considered system by solving the formulated problem in the infinite SNR regime.
Section \ref{sec:Proposed Solution at Finite SNR} proposes an efficient algorithm to solve the formulated problem in the finite SNR regime. Section \ref{sec:Numerical Examples} provides the numerical simulation results to evaluate the performance of the proposed strategy. Finally, Section \ref{sec:Conclusion} concludes the paper.

\section{System Model}\label{sec:System Model}
\begin{figure}[t]
  \centering
  \includegraphics[width=7cm]{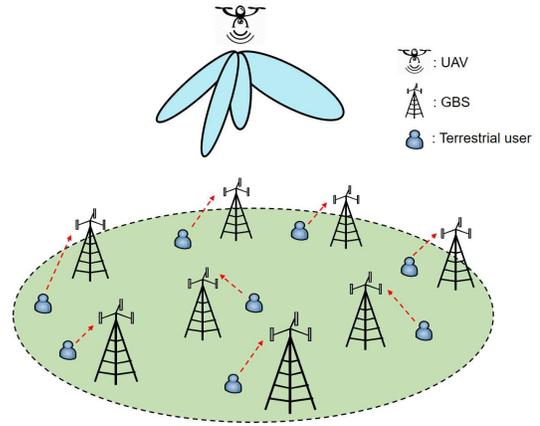}
  \caption{Illustration of multi-beam UAV communication in the cellular uplink.}\label{system_model}
\end{figure}
As shown in Fig. \ref{system_model}, we consider the uplink communication in a cellular network by a UAV equipped with $M>1$ antennas, while there are $N$ GBSs in the UAV's signal coverage region denoted by the set $\mathcal{N}=\{1,\cdots,N\}$. It is assumed that each GBS $n\in \mathcal{N}$ already serves one terrestrial user in the same RB as that assigned to the UAV. In this paper, we focus on the challenging case when the number of (occupied) GBSs is larger than the number of antennas at the UAV, i.e., $N>M$. This is usually the case in practice due to the macro-diversity and the high frequency reuse factor in today's cellular networks as well as the space and hardware limitations to install multiple antennas at the UAV.

We assume that each GBS has a fixed beam pattern for the aerial user, thus can be equivalently viewed as being equipped with one single antenna for the purpose of exposition. The effective channel from the UAV to GBS $n$ is thus an $M$ by $1$ vector denoted by $\mv{h}_n\in \mathbb{C}^{M\times 1}$. In this paper, we consider that the UAV is at a given location in the air, where $d_n$ in meters (m) denotes the distance from the UAV to GBS $n$. Moreover, we consider the practical Rician fading channel model, and the channel from the UAV to GBS $n$ is given by
\begin{align}\label{eqn:Rician channel}
\mv{h}_n=\sqrt{\frac{\tau_0}{d_n^2}}\left(\sqrt{\frac{\lambda}{\lambda+1}}\hat{\mv{h}}_n+\sqrt{\frac{1}{\lambda+1}}\tilde{\mv{h}}_n\right), ~~~ \forall n,
\end{align}where $\tau_0$ denotes the channel power gain at the reference distance $d_0=1$ m; $\hat{\mv{h}}_n\in \mathbb{C}^{M\times 1}$ denotes the LoS channel component; $\tilde{\mv{h}}_n\in \mathbb{C}^{M\times 1}$ with $\tilde{\mv{h}}_n\sim \mathcal{CN}(\mv{0},\mv{I})$ denotes the Rayleigh fading channel component; and $\lambda\geq 0$ is the Rician factor specifying the power ratio between the LoS and Rayleigh fading components in $\mv{h}_n$. Note that under the above Rician fading channel model, the channel vectors between the multi-antenna UAV and any $N',\ N'\leq \min(M,N)=M$, GBSs are linearly independent with probability one. We further assume that all ${\mv{h}}_n$'s are known at the UAV.\footnote{In practice, the UAV may send $M$ orthogonal pilots for channel training, and the GBSs then feed back their estimated channels to the UAV.}

The UAV aims to transmit its messages to the GBSs via spatial multiplexing with transmit beamforming. Specifically, its transmit signal is expressed as
\begin{align}\label{eqn:transmit signal}
\mv{x}=\sum\limits_{j=1}^J\mv{w}_js_j,
\end{align}where $J\leq \min(M,N)=M$ denotes the number of data streams sent by the UAV, $s_j\sim \mathcal{CN}(0,1)$ denotes the message of the $j$th data stream, and $\mv{w}_j \in \mathbb{C}^{M\times 1}$ denotes the corresponding beamforming vector. The received signal at the $n$th GBS is given by
\begin{align}
y_n&=\mv{h}_n^H\mv{x}+S_n+z_n \nonumber \\
&=\mv{h}_n^H\sum\limits_{j=1}^J\mv{w}_js_j+S_n+z_n, ~~~ \forall n, \label{eqn:received signal}
\end{align}where we assume $S_n\sim \mathcal{CN}(0,Q_n)$ denotes the received signal from the terrestrial user served by the $n$th GBS, and $z_n\sim \mathcal{CN}(0,\sigma_n^2)$ denotes the superposition of the additive white Gaussian noise (AWGN) at the $n$th GBS and the interference from the terrestrial users in the other cells. Note that each single-antenna GBS $n$ can decode at most one data stream sent by the UAV.

\section{Multi-Beam Transmission with NOMA}\label{sec:NOMA-based Strategy}

There are two main goals in designing the UAV multi-beam transmission: first, each UAV data stream should be decoded by at least one GBS; and second, the rates of the terrestrial users cannot be degraded by the UAV uplink communications, i.e., the interference generated from the UAV to each GBS $n$ for decoding its served terrestrial user's message in $S_n$ needs to be nulled.

To fulfill the first goal, we define $\Lambda_j\subseteq \mathcal{N}$ as the set of GBSs that decode the $j$th data stream sent by the UAV, i.e., $s_j$, $j\in \{1,\cdots,J\}$. Since each data stream has to be decoded by at least one GBS, and each GBS can decode at most one data stream, it follows that
\begin{align}
& \Lambda_j\neq \emptyset, ~~~ \forall j, \label{eqn:no empty} \\
& \Lambda_j\bigcap \Lambda_i =\emptyset, ~~~ \forall j, ~ i\neq j. \label{eqn:one data stream}
\end{align}

According to (\ref{eqn:received signal}), if the $j$th data stream is decoded by the $n$th GBS, i.e., $n\in \Lambda_j$, its decoding signal-to-interference-plus-noise ratio (SINR) is
\begin{align}\label{eqn:SINR}
\gamma_{j,n}=\frac{|\mv{h}_n^H\mv{w}_j|^2}{\sum\limits_{i\neq j}|\mv{h}_n^H\mv{w}_i|^2+Q_n+\sigma_n^2}, ~~~ \forall n\in \Lambda_j.
\end{align}Then, the following inequality needs to hold:
\begin{align}\label{eqn:rate}
\log_2\left(1+\gamma_{j,n}\right)\geq R_j, ~~~ \forall j, ~ n\in \Lambda_j,
\end{align}where $R_j$ in bits/second/Hz is the transmit rate of data stream $j$.

Next, consider the second goal of interference avoidance to the terrestrial communications. According to (\ref{eqn:received signal}), the power of the received interference at GBS $n$ from the UAV is
\begin{align}\label{eqn:interference 1}
I_n=\sum\limits_{j=1}^J|\mv{h}_n^H\mv{w}_j|^2, ~~~ \forall n.
\end{align}Thus, the interference avoidance constraints are expressed as
\begin{align}\label{eqn:interference temperature 1}
\sum\limits_{j=1}^J|\mv{h}_n^H\mv{w}_j|^2=0, ~~~ \forall n.
\end{align}However, if any GBS, say GBS $n$, decodes $s_j$, i.e., $n\in \Lambda_j$, it must follow that $|\mv{h}_n^H\mv{w}_j|^2>0$ to achieve a positive transmit rate as given in (\ref{eqn:rate}), which already violates constraint (\ref{eqn:interference temperature 1}). As a result, by simply treating UAV interference as noise, the UAV cannot transmit any data stream without degrading the rates of the terrestrial users.

This motivates us to apply NOMA jointly with ZF beamforming to guarantee a positive transmit rate of UAV while at the same time satisfying the interference avoidance constraints for the terrestrial communications. Specifically each GBS that needs to decode one data stream from the UAV first decodes the UAV message, then subtracts it from its received signal, and at last decodes the message of its associated terrestrial user. The details of the proposed strategy are given as follows.

For convenience, define $\Phi_n=\{j:n\in \Lambda_j\}$ as the set of data streams that are decoded by GBS $n$. Since each GBS can decode at most one data stream, we have $|\Phi_n|\leq 1$, $\forall n$. If GBS $n$ does not decode any message from the UAV, i.e., $\Phi_n=\emptyset$, then according to (\ref{eqn:received signal}), the power of the interference from the UAV to GBS $n$ is the same as that given in (\ref{eqn:interference temperature 1}). However, if GBS $n$ decodes $s_j$, i.e., $\Phi_n=\{j\}$, then it can cancel the interference caused by $s_j$ before decoding the message of its served terrestrial user. Specifically, after interference cancellation, the received signal given in (\ref{eqn:received signal}) reduces to
\begin{align}\label{eqn:received signal 1}
y_n=S_n+\mv{h}_n^H\sum\limits_{i\neq j}\mv{w}_is_i+z_n, ~~~ {\rm if} ~ \Phi_n=\{j\}.
\end{align}As a result, the power of the (residue) interference from the UAV to GBS $n$ becomes
\begin{align}\label{eqn:interference 2}
I_n=\sum\limits_{i\neq j}|\mv{h}_n^H\mv{w}_i|^2, ~~~ {\rm if} ~ \Phi_n=\{j\}.
\end{align}

To summarize, under the proposed NOMA-based strategy, the new ZF constraint for interference mitigation is expressed as
\begin{align}\label{eqn:interference temperature}
& \sum\limits_{j\notin \Phi_n}|\mv{h}_n^H\mv{w}_j|^2=0, ~~~ \forall n.
\end{align}

In the following, we provide a simple example to shed some light on the effectiveness of the proposed NOMA-based strategy in our considered system.

\begin{example}\label{example1}
Consider the case when the UAV has $M=3$ antennas, and there are $N=4$ GBSs. Notice that with traditional ZF beamforming only, no data stream can be transmitted without generating any interference to the GBSs. In contrast, under our proposed NOMA-based strategy, the UAV can send $J=2$ data streams to the GBSs as follows: data stream 1 is sent to GBSs 1 and 2, while data stream 2 is sent to GBSs 3 and 4, i.e., $\Lambda_1=\{1,2\}$ and $\Lambda_2=\{3,4\}$. In this case, since the UAV has 3 antennas, it is feasible to align $\mv{w}_1$ to the null space of $\mv{h}_3$ and $\mv{h}_4$, i.e., $\mv{h}_3^H\mv{w}_1=\mv{h}_4^H\mv{w}_1=0$, and align $\mv{w}_2$ to the null space of $\mv{h}_1$ and $\mv{h}_2$, i.e., $\mv{h}_1^H\mv{w}_2=\mv{h}_2^H\mv{w}_2=0$, such that data streams 1 and 2 do not generate interference at GBSs 3, 4 and GBSs 1, 2, respectively. Moreover, with NOMA, the interference from data streams 1 and 2 is cancelled at GBSs 1, 2 and GBSs 3, 4, respectively. As a result, the interference from the UAV to the terrestrial communications is effectively zero at all the 4 GBSs.
\end{example}

\section{Problem Formulation}\label{sec:Problem Formulation}
Under the NOMA-based strategy proposed in Section \ref{sec:NOMA-based Strategy}, we aim to maximize the transmit rate of the UAV to the GBSs subject to the interference constraints (\ref{eqn:interference temperature}). To this end, we jointly optimize the number of data streams sent by the UAV, i.e., $J$, the data stream association, i.e., $\Lambda_j$'s, the transmit beamforming vectors, i.e., $\mv{w}_j$'s, and the transmit rates of various data streams, i.e., $R_j$'s. Specifically, our problem of interest is formulated as
\begin{subequations}\label{eqn:P1}\begin{align}
\mathop{ \underset{J,\{\Lambda_j,\mv{w}_j,R_j\}}{\max}} & ~ \sum\limits_{j=1}^J  R_j \label{eqn:problem}\\
\mathrm{s.t.} \ \ \ \ & (\ref{eqn:no empty}), ~ (\ref{eqn:one data stream}), ~ (\ref{eqn:rate}), ~ (\ref{eqn:interference temperature}), \\ & \sum\limits_{j=1}^J\|\mv{w}_j\|^2 \leq P, \label{eqn:power constraint}
\end{align}\end{subequations}where $P$ is the maximum transmit power of the UAV.

Without the zero-interference constraints (\ref{eqn:interference temperature}) for the terrestrial users, the optimal data stream association to the above problem must satisfy $|\Lambda_j|=1$, $\forall j$, since sending one data stream to more than one GBSs may degrade its rate. However, with the zero-interference constraints (\ref{eqn:interference temperature}) in problem (\ref{eqn:P1}), in general we need to send one data stream to multiple GBSs, i.e., $|\Lambda_j|>1$ for some $j$'s, with certain rate compromise, such that more GBSs can cancel the interference caused by this data stream before decoding the messages from their served terrestrial users. The key challenge for solving problem (\ref{eqn:P1}) thus lies in how to determine the data stream association jointly with transmit beamforming: for each data stream, should we align its beamforming vector in the null space of the channel of a GBS for interference nulling, or in the same direction of its channel for interference cancellation? To tackle the above challenge and gain essential insight, we first characterize the DoF achievable for sum-rate maximization by solving problem (\ref{eqn:P1}) at the infinite SNR regime, i.e, $P\rightarrow \infty$, in Section \ref{sec:DoF Analysis}, and then propose an efficient algorithm to solve problem (\ref{eqn:P1}) at finite SNR regime in Section \ref{sec:Proposed Solution at Finite SNR}.

\section{DoF Analysis}\label{sec:DoF Analysis}
Note that in the high-SNR regime, the DoF is a fundamental characterization of the achievable rate in multi-antenna communication systems \cite{Jafar08}. Due to the strong LoS channels between the UAV and GBSs, the high-SNR assumption is practically valid in our considered system. As a result, in this section we study the maximum DoF of the multi-beam UAV uplink communication under our proposed NOMA-based strategy. The obtained design that maximizes the DoF will also be useful to our proposed solution for problem (\ref{eqn:P1}) in the finite SNR regime in Section \ref{sec:Proposed Solution at Finite SNR}.

The DoF represents the rate of growth for the system capacity with respect to the logarithm of the SNR \cite{Jafar08}. In our considered system, a DoF of $J$ is achievable if $J$ data streams can be transmitted from the UAV such that each data stream $j$, $1\leq j \leq J$, can be decoded by at least one GBS, while its interference at all GBSs needs to be zero. Specifically, zero interference of $s_j$ by each data stream at any GBS can be achieved by either aligning $\mv{w}_j$ to the null space of this GBS's channel or otherwise such that this GBS can decode $s_j$ for interference cancellation. Mathematically, the maximum achievable DoF can be obtained by solving problem (\ref{eqn:P1}) with $P\rightarrow \infty$, i.e.,
\begin{subequations}\label{eqn:DoF}\begin{align}
\mathop{ \underset{J,\{\Lambda_j,\mv{w}_j\}}{\max}} & ~ J \label{eqn:problem 1}\\
\mathrm{s.t.} \ \ & (\ref{eqn:no empty}), ~ (\ref{eqn:one data stream}), \\
& \mv{h}_n^H\mv{w}_j\neq 0, ~~~ \forall n\in \Lambda_j, ~ \forall j, \label{eqn:DoF signal} \\
& \mv{h}_n^H\mv{w}_j=0, ~~~ \forall n\notin \Lambda_j, ~ \forall j. \label{eqn:linear equations}
\end{align}\end{subequations}In problem (\ref{eqn:DoF}), constraint (\ref{eqn:DoF signal}) guarantees that each data stream can be decoded by its associated GBSs, and constraint (\ref{eqn:linear equations}), which is equivalent to constraint (\ref{eqn:interference temperature}), ensures that each data stream does not generate any interference to GBSs that do not decode this data stream.

\begin{theorem}\label{Theorem1}
Under the Rician fading channel model (\ref{eqn:Rician channel}) and assuming $N>M$, the maximum DoF of the considered system under the proposed NOMA-based strategy, i.e., the optimal value of problem (\ref{eqn:DoF}), is given by
\begin{align}\label{eqn:maximum DoF}
J^\ast=\left\lfloor \frac{N}{N-M+1} \right \rfloor,
\end{align}where $\lfloor x \rfloor$ denotes the maximum integer no larger than $x$.
\end{theorem}

\begin{IEEEproof}
Please refer to Appendix \ref{appendix1}.
\end{IEEEproof}

Note that if $N=4$ and $M=3$, the maximum DoF is $J^\ast=2$ according to Theorem \ref{Theorem1}, as shown in Example \ref{example1}.

Some intuition behind Theorem \ref{Theorem1} is as follows. First, according to (\ref{eqn:maximum DoF}), we always have $J^\ast\geq 1$. In other words, at least one data stream can be transmitted by the UAV without generating any interference to the terrestrial communications. Note that without NOMA, no data stream can be transmitted by the UAV with the zero-interference constraints at all GBSs under $N>M$.

Second, under $N>M$, it can be shown that $J^\ast$ given in (\ref{eqn:maximum DoF}) must satisfy $J^\ast<M$. As a result, the upper bound of the DoF is $M$, which is the maximum DoF when all the GBSs are fully connected by the backhaul links such that a joint information decoding is feasible. The achievable DoF under our strategy shown in Theorem \ref{Theorem1} is usually a fraction of this upper bound. For example, if $N=M+1$, then the maximum DoF is $J^\ast=\lfloor (M+1)/2 \rfloor$, which is about half of the DoF achieved in the ideal case with fully connected GBSs.

Last, as shown in Appendix \ref{appendix1}, to achieve the maximum DoF given in (\ref{eqn:maximum DoF}), the data stream-GBS associations should satisfy
\begin{align}
& |\Lambda_j|=\left\lfloor \frac{N}{J^\ast} \right\rfloor, ~~~ j=1,\cdots,j^\ast, \label{eqn:data association 1} \\
& |\Lambda_j|=\left\lceil \frac{N}{J^\ast} \right\rceil, ~~~ j=j^\ast+1,\cdots,J^\ast, \label{eqn:data association 2}
\end{align}where $\lceil x \rceil$ denotes the minimum integer no smaller than $x$, and $j^\ast\in [1,J^\ast]$ is the unique integer such that $\sum_j |\Lambda_j|=N$. It can be shown that if the data stream-GBS associations satisfy (\ref{eqn:data association 1}) and (\ref{eqn:data association 2}), we can always construct feasible ZF beamforming vectors satisfying the zero-interference constraint (\ref{eqn:linear equations}) since $M>N-|\Lambda_j|$, $\forall j$.

\section{Proposed Solution at Finite SNR}\label{sec:Proposed Solution at Finite SNR}

In this section, we propose an efficient algorithm to solve problem (\ref{eqn:P1}) sub-optimally based on the results in the last section. Note that given $J$ and $\Lambda_j$'s, problem (\ref{eqn:P1}) reduces to a beamforming design problem in a multi-group multicast system, for which suboptimal algorithms have been designed in \cite{Liang18}. However, it is difficult to jointly optimize $J$, $\Lambda_j$'s, $\mv{w}_j$'s and $R_j$'s. In this section, we propose a method to decouple the design of $J$ and $\Lambda_j$'s from that of $\mv{w}_j$'s and $R_j$'s. First, we design $J$ and $\Lambda_j$'s based on the optimal solution to problem (\ref{eqn:DoF}), since problem (\ref{eqn:DoF}) is a good approximation of problem (\ref{eqn:P1}) in the high-SNR regime due to the strong LoS channels between the UAV and GBSs. Then, with the obtained $J$ and $\Lambda_j$'s, we design the beamforming vectors to maximize the UAV sum-rate.

First, we set $J=J^\ast$ in problem (\ref{eqn:P1}), and determine the cardinalities of $\Lambda_j$'s based on (\ref{eqn:data association 1}) and (\ref{eqn:data association 2}). The remaining question is how to assign $|\Lambda_j|$ elements to each set $\Lambda_j$. Note that with the zero-interference constraint (\ref{eqn:interference temperature}), the rate constraint (\ref{eqn:rate}) in problem (\ref{eqn:P1}) reduces to
\begin{align}\label{eqn:rate 2}
\log_2\left(1+\frac{|\mv{h}_n^H\mv{w}_j|^2}{Q_n+\sigma_n^2}\right)\geq R_j, ~~~ \forall n\in \Lambda_j, ~ \forall j.
\end{align}Intuitively, the rate for multicasting $s_j$ critically depends on the GBS with the lowest effective SINR defined as $\|\mv{h}_n\|^2/(Q_n+\sigma_n^2)$ in the set $\Lambda_j$. If in each set $\Lambda_j$, there exists a GBS with very poor effective SINR, then the multicasting rates are small for all the data streams. Motivated by this observation, we propose to multicast each data stream to a set of GBSs with similar effective SINRs. Specifically, we re-arrange the order of GBSs as $n_1,\cdots,n_N$ such that $\|\mv{h}_{n_1}\|^2/(Q_{n_1}+\sigma_{n_1}^2) \geq \|\mv{h}_{n_2}\|^2/(Q_{n_2}+\sigma_{n_2}^2) \geq \cdots \geq \|\mv{h}_{n_N}\|^2/(Q_{n_N}+\sigma_{n_N}^2)$. Then, we adopt the following data stream association solution:
\begin{align}\label{eqn:data stream association solution}
\Lambda_j=\{n_{\sum_{i<j}|\Lambda_i|+1},\cdots,n_{\sum_{i\leq j}|\Lambda_i|}\}, ~ j=1,\cdots,J^\ast,
\end{align}where $|\Lambda_j|$'s are given in (\ref{eqn:data association 1}) and (\ref{eqn:data association 2}).

Next, given $J=J^\ast$ and $\Lambda_j$'s as given in (\ref{eqn:data stream association solution}), problem (\ref{eqn:P1}) reduces to the following beamforming design problem:
\begin{subequations}\label{eqn:P2}\begin{align}
\mathop{ \underset{\{\mv{w}_j,R_j\}}{\max}} & ~ \sum\limits_{j=1}^J  R_j \label{eqn:problem 2}\\
\mathrm{s.t.} \ \  & (\ref{eqn:interference temperature}), ~ ({\rm \ref{eqn:power constraint}}), ~ (\ref{eqn:rate 2}).
\end{align}\end{subequations}In the following, we reduce the complexity for solving the above problem based on the zero-interference constraints (\ref{eqn:interference temperature}). For simplicity, define
\begin{align}
\mv{H}_{-j}=[\cdots,\mv{h}_n,\cdots,\forall n\notin \Lambda_j]^H\in \mathbb{C}^{(N-|\Lambda_j|)\times M}, ~ \forall j, \end{align}as the collection of channels of all the GBSs that do not decode the $j$th data stream of the UAV. As a result, the zero-interference constraint (\ref{eqn:interference temperature}) is equivalent to
\begin{align}
\bar{\mv{H}}_{-j}\mv{w}_j=0, ~~~ \forall j.
\end{align}

Define the singular value decomposition (SVD) of $\bar{\mv{H}}_{-j}$ as
\begin{align}
\bar{\mv{H}}_{-j}=\mv{X}_j\mv{\Theta}_j\mv{Y}_j^H=\mv{X}_j\mv{\Theta}_j[\bar{\mv{Y}}_j,\tilde{\mv{Y}}_j]^H, ~~~ \forall j,
\end{align}where $\mv{X}_j\in \mathbb{C}^{(N-|\Lambda_j|)\times (N-|\Lambda_j|)}$ and $\mv{Y}_j\in \mathbb{C}^{M\times M}$ are unitary matrices, and $\mv{\Theta}_k$ is a $(N-|\Lambda_j|)\times M$ rectangular diagonal matrix. Furthermore, $\bar{\mv{Y}}_j\in \mathbb{C}^{M\times (N-|\Lambda_j|)}$ and $\tilde{\mv{Y}}_j\in \mathbb{C}^{M\times (M-(N-|\Lambda_j|))}$ consist of the first $N-|\Lambda_j|$ and the last $M-(N-|\Lambda_j|)$ right singular vectors of $\bar{\mv{H}}_{-j}$, respectively. Note that $\tilde{\mv{Y}}_j$ forms an orthogonal basis for the null space of $\bar{\mv{H}}_{-j}$, thus $\mv{w}_j$ must be in the following form:
\begin{align}
\mv{w}_j=\tilde{\mv{Y}}_j\tilde{\mv{w}}_j, ~~~ \forall j,
\end{align}where $\tilde{\mv{w}}_j\in \mathbb{C}^{(M-(N-|\Lambda_j|))\times 1}$. As a result, with the above ZF beamforming solution, problem (\ref{eqn:P2}) is equivalent to
\begin{subequations}\label{eqn:P3}\begin{align}
\mathop{ \underset{\{\tilde{\mv{w}}_j,R_j\}}{\max}} & ~ \sum\limits_{j=1}^J  R_j \label{eqn:problem 3}\\
\mathrm{s.t.} \ \  & \log_2\left(1+\frac{|\mv{h}_n^H\tilde{\mv{Y}}_j\tilde{\mv{w}}_j|^2}{Q_n+\sigma_n^2}\right)\geq R_j, ~ \forall j, ~ \forall j\in \Lambda_j, \\ & \sum\limits_{j=1}^J\|\tilde{\mv{Y}}_j\tilde{\mv{w}}_j\|^2\leq P.
\end{align}\end{subequations}As compared to problem (\ref{eqn:P2}), the number of variables for designing each beamforming vector is reduced from $M$ in $\mv{w}_j$'s to $M-(N-|\Lambda_j|)$ in $\tilde{\mv{w}}_j$'s. Moreover, the number of constraints is also reduced since the zero-interference constraints (\ref{eqn:interference temperature}) do not need to be expressed explicitly. Note that problem (\ref{eqn:P3}) can be efficiently solved similarly by Algorithm 1 in \cite{Liang18}. We thus omit the details here.

\section{Numerical Examples}\label{sec:Numerical Examples}
\begin{figure}[t]
  \centering
  \includegraphics[width=8cm]{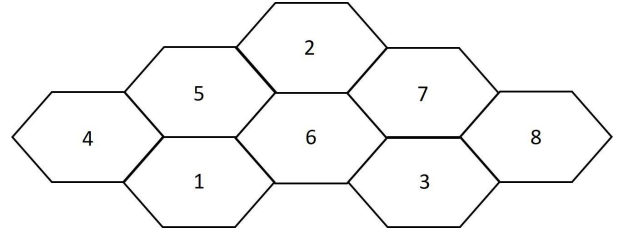}
  \caption{Cellular network topology for simulation.}\label{setup}\vspace{-5pt}
\end{figure}
In this section, we provide numerical examples to verify the effectiveness of the proposed multi-beam transmission strategy with NOMA. We assume that the UAV is at a height of $H=100$ m, and $N=8$ GBSs are in the UAV's signal coverage area, as depicted in Fig. \ref{setup}. Moreover, it is assumed that one terrestrial user is randomly located in each cell, which communicates with the GBS using the same RB as the UAV. The transmit power of each terrestrial user is assumed to be $23$ dBm. We further assume that in the Rician fading channel model (\ref{eqn:Rician channel}), the Rician factor is $\lambda=3$, and the LoS components $\hat{\mv{h}}_n$'s follow a linear array model \cite{Luo07}. At last, the bandwidth of the RB used by the UAV is $10$ MHz, while the power spectrum density of the AWGN at the GBSs is $-169$ dBm.

\subsection{DoF Performance}
First, we consider the DoF performance achieved by our proposed multi-beam transmission strategy with NOMA. In the following, we list the maximum DoF for various values of $M$ under $M<N$ according to Theorem \ref{Theorem1} and the corresponding DoF-achievable data stream association solution.

\begin{enumerate}
\item $M=1,2,3,4$: Maximum DoF $J^\ast=1$ is achievable with $\Lambda_1=\{1,\cdots,8\}$.
\item $M=5,6$: Maximum DoF $J^\ast=2$ is achievable with $\Lambda_1=\{1,2,3,4\}$ and $\Lambda_2=\{5,6,7,8\}$.
\item $M=7$: Maximum DoF $J^\ast=4$ is achievable with $\Lambda_1=\{1,2\}$, $\Lambda_2=\{3,4\}$, $\Lambda_3=\{5,6\}$, and $\Lambda_4=\{7,8\}$.
\end{enumerate}

For comparison, we evaluate an upper bound and a lower bound of the maximum DoF in our considered system. Specifically, for the performance upper bound, we consider the case that all the GBSs are connected to the same central processor, where a joint information decoding can be done based on the signals received by all the GBSs. In this case, the maximum DoF is $J=\min(M,N)=M$. For the performance lower bound, we consider the case that NOMA is not applied for interference cancellation. As explained in Section \ref{sec:NOMA-based Strategy}, the DoF is always $0$ under our considered setup with $N>M$.

\begin{figure}[t]
  \centering
  \includegraphics[width=8cm]{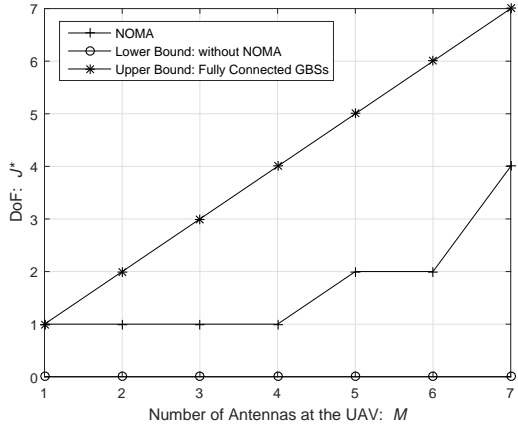}
  \caption{Maximum achievable DoF versus number of antennas at the UAV.}\label{DoF}\vspace{-5pt}
\end{figure}

The maximum DoF achieved by our proposed multi-beam transmission strategy with NOMA as well as the DoF upper bound and lower bound are depicted in Fig. \ref{DoF}, when the UAV is equipped with $M=1,\cdots,7$ antennas. It is observed that as compared to the DoF upper bound, our proposed strategy can achieve about half of the DoF for various values of $M$. However, our proposed strategy is more practically appealing since it merely relies on NOMA for interference cancellation, rather than requiring bachkaul links for centralized interference cancellation. Moreover, thanks to NOMA, our proposed strategy achieves significant DoF gain as compared to the DoF lower bound for various values of $M$.
\subsection{Transmit Rate Performance}
Next, we consider the transmit rate performance of the UAV at finite SNR regime. It is assumed that the UAV is equipped with $M=6$ antennas. Since the maximum DoF is $J^\ast=2$ as shown in Fig. \ref{DoF}, we assume $J=2$ in problem (\ref{eqn:P1}). Moreover, according to (\ref{eqn:data stream association solution}), we adopt the following data stream association solution in problem (\ref{eqn:P1}): $\Lambda_1=\{2,3,5,6\}$ and $\Lambda_2=\{1,4,7,8\}$. At last, we use Algorithm 1 in \cite{Liang18} to solve problem (\ref{eqn:P2}) for finding the beamforming solution to problem (\ref{eqn:P1}). Note that the key to solving problem (\ref{eqn:P1}) lies in the data stream association strategy. To illustrate the effectiveness of the proposed data stream association solution (\ref{eqn:data stream association solution}), we also consider a benchmark scheme in which $4$ GBSs are randomly selected for decoding the first data stream of the UAV, while the other $4$ GBSs decode the second data stream.

\begin{figure}[t]
  \centering
  \includegraphics[width=8cm]{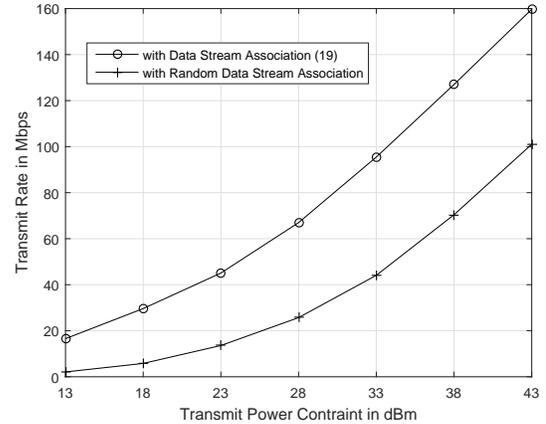}
  \caption{Transmit rate of UAV versus its transmit power constraint.}\label{rate}\vspace{-5pt}
\end{figure}

Fig. \ref{rate} shows the transmit rate of the UAV when its transmit power constraint $P$ varies from $13$ dBm to $43$ dBm, for both the cases when the data stream association solution is determined based on (\ref{eqn:data stream association solution}) or randomly selected. For simplicity, we assume that in the benchmark scheme, the random data stream association solution is $\Lambda_1=\{1,2,3,4\}$ and $\Lambda_2=\{5,6,7,8\}$. It is observed that for both the cases, the transmit rate of the UAV increases with its transmit power constraint. Moreover, it is observed that for various values of the transmit power constraint, the transmit rate of the UAV achieved by the data stream association solution based on (\ref{eqn:data stream association solution}) is much higher than that achieved by the benchmark scheme with random data stream association. This is because under the considered setup, GBSs $4$ and $7$ have the lowest effective SINR among all the GBSs. With the solution $\Lambda_1=\{2,3,5,6\}$ and $\Lambda_2=\{1,4,7,8\}$, the transmit rate of the second data stream is limited by GBSs $4$ and $7$ and thus practically low, but the transmit rate of the first data stream is much higher. However, with the solution $\Lambda_1=\{1,2,3,4\}$ and $\Lambda_2=\{5,6,7,8\}$, the transmit rates of the first and second data streams are limited by GBSs $4$ and $7$, respectively, and are thus both relatively low. As a result, our proposed data stream association solution is more practically effective.

\section{Conclusion}\label{sec:Conclusion}
This paper studied the UAV uplink communication in the cellular network, where a multi-antenna UAV sends multiple data streams to a large number of GBSs each serving a terrestrial user at the same time over the same frequency band. To achieve high data rate while mitigating the severe interference to the terrestrial communications due to the strong UAV-GBS LoS channels, this paper proposed a novel NOMA-based transmission strategy where a selected subset of the GBSs first decode the UAV's signals and then cancel them for decoding the terrestrial users' signals. The achievable DoF for UAV's sum-rate maximization was first characterized in closed-form in the infinite SNR regime, based on which an efficient algorithm was then proposed to maximize the UAV's sum-rate in the finite SNR regime, both subjected to the zero-interference constraints for protecting the terrestrial communications. Numerical examples were provided to verify the effectiveness of the proposed strategy and optimized design, which was shown able to significantly enhance the UAV's uplink throughput in the cellular network.

\begin{appendix}
\subsection{Proof of Theorem \ref{Theorem1}}\label{appendix1}
First, we show that given any $J\leq J^\ast$, we can always find a set of solution $\Lambda_j$'s and $\mv{w}_j$'s such that all the constraints are satisfied in problem (\ref{eqn:DoF}). In other words, the optimal value of problem (\ref{eqn:DoF}) is no smaller than $J^\ast$. Given any $J<N$, we can always construct $\Lambda_1,\cdots,\Lambda_J$ that satisfy (\ref{eqn:no empty}), (\ref{eqn:one data stream}), $|\Lambda_j|=\lfloor N/J \rfloor$ for $j=1,\cdots,j^\ast$, $|\Lambda_j|=\lceil N/J \rceil$ for $j=j^\ast+1,\cdots,J$, and $\sum_j|\Lambda_j|=N$, for some $j^\ast \in [1,J]$. For each data stream $j$, the corresponding beamforming $\mv{w}_j$ should satisfy (\ref{eqn:linear equations}), which is a linear system with $M$ variables and $N-|\Lambda_j|$ linear equations. Note that with the above construction of $\Lambda_j$'s, we have
\begin{align}\label{eqn:dimension}
N- |\Lambda_j|\leq N-\left\lfloor \frac{N}{J} \right\rfloor, ~~~ \forall j.
\end{align}If $J\leq J^\ast$, it can be shown that
\begin{align}
\left\lfloor \frac{N}{J} \right\rfloor \geq \left\lfloor \frac{N}{J^\ast} \right\rfloor = N-M+1>N-M,
\end{align}or equivalently $M>N-\lfloor N/J \rfloor\geq N-|\Lambda_j|$. In other words, the number of variables is larger than that of linear equations, and thus there always exist a non-zero solution of $\mv{w}_j$ that satisfies (\ref{eqn:linear equations}), $\forall j$. Moreover, under the Rician fading channel model (\ref{eqn:Rician channel}), all the channels $\mv{h}_n$'s are independent with each other. As a result, if $\mv{w}_j$ is designed based on $\mv{h}_n$'s, $\forall n\in \cup_{i\neq j} \Lambda_i$, as shown in (\ref{eqn:linear equations}), then with probability one we have $\mv{h}_n^H\mv{w}_j\neq 0$, $\forall n\in \Lambda_j$, i.e., (\ref{eqn:DoF signal}). To summarize, given any $J\leq J^\ast$, we are able to construct a feasible solution to problem (\ref{eqn:DoF}).

Next, we show that if $J>J^\ast$, then we can never find a set of solution $\Lambda_j$'s and $\mv{w}_j$'s such that all the constraints are satisfied in problem (\ref{eqn:DoF}). In other words, the optimal value of problem (\ref{eqn:DoF}) cannot be larger than $J^\ast$. Given any solution $\Lambda_j$'s, we need to design $\mv{w}_j$'s to satisfy the linear equations shown in (\ref{eqn:linear equations}). Note that given any $J$, there must exists a $j^\ast\in [1,J]$ such that $|\Lambda_{j^\ast}|\leq \lfloor N/J \rfloor$, since otherwise we have $\sum_j|\Lambda_j|> N$, which cannot be true under the constraint (\ref{eqn:one data stream}). For this particular $j^\ast$, the number of linear equations as in (\ref{eqn:linear equations}) satisfies
\begin{align}\label{eqn:dimension 2}
N- |\Lambda_{j^\ast}|\geq N-\left\lfloor \frac{N}{J} \right\rfloor.
\end{align}If $J>J^\ast$, it can be shown that
\begin{align}
\left\lfloor \frac{N}{J} \right\rfloor < \left\lfloor \frac{N}{J^\ast} \right\rfloor =N-M+1,
\end{align}or equivalently $M\leq N-\lfloor N/J \rfloor\leq N-|\Lambda_{j^\ast}|$. In other words, the number of variables is no larger than that of linear equations. Moreover, in the linear equations shown in (\ref{eqn:linear equations}), $\mv{h}_n$'s are independent with each other under the Rician fading channel model (\ref{eqn:Rician channel}). As a result, with probability one, we cannot find a non-zero $\mv{w}_{j^\ast}$ that satisfies $\mv{h}_n^H\mv{w}_{j^\ast}=0$, $\forall n\in \cup_{i\neq j^\ast}\Lambda_i$.

As a result, the maximum DoF, i.e., the optimal value of problem (\ref{eqn:DoF}), is equal to $J^\ast$. Theorem \ref{Theorem1} is thus proved.
\end{appendix}

\bibliographystyle{IEEEtran}
\bibliography{CIC}

\begin{thebibliography}{10}
\providecommand{\url}[1]{#1}
\csname url@samestyle\endcsname
\providecommand{\newblock}{\relax}
\providecommand{\bibinfo}[2]{#2}
\providecommand{\BIBentrySTDinterwordspacing}{\spaceskip=0pt\relax}
\providecommand{\BIBentryALTinterwordstretchfactor}{4}
\providecommand{\BIBentryALTinterwordspacing}{\spaceskip=\fontdimen2\font plus
\BIBentryALTinterwordstretchfactor\fontdimen3\font minus
  \fontdimen4\font\relax}
\providecommand{\BIBforeignlanguage}[2]{{%
\expandafter\ifx\csname l@#1\endcsname\relax
\typeout{** WARNING: IEEEtran.bst: No hyphenation pattern has been}%
\typeout{** loaded for the language `#1'. Using the pattern for}%
\typeout{** the default language instead.}%
\else
\language=\csname l@#1\endcsname
\fi
#2}}
\providecommand{\BIBdecl}{\relax}
\BIBdecl

\bibitem{Zhang18}
S.~Zhang, Y.~Zeng, and R.~Zhang, ``{Cellular-enabled UAV communication: A
  connectivity-constrained trajectory optimization perspective},'' submitted to
  \emph{IEEE Trans. Commun.} [Online] Available:
  https://arxiv.org/abs/1805.07182.

\bibitem{LTE}
B.~V.~D. Bergh, A.~Chiumento, and S.~Pollin, ``{LTE in the sky: Trading off
  propagation benefits with interference costs for aerial nodes},'' \emph{IEEE
  Commun. Mag.}, vol.~54, no.~5, pp. 44--50, May 2016.

\bibitem{CellularConnected}
Y.~Zeng, J.~Lyu, and R.~Zhang, ``{Cellular-connected UAVs: Potentials,
  challenges and promising technologies},'' to appear in \emph{IEEE Wireless
  Commun.} [Online] Available: https://arxiv.org/abs/1804.02217.

\bibitem{Qualcomm}
``{LTE unmanned aircraft systems},'' May 2017, {Qualcomm Technologies Inc.,
  Trial} {Report} v1.0.1.

\bibitem{Azari17}
M.~Azari, F.~Rosas, A.~Chiumento, and S.~Pollin, ``{Coexistence of terrestrial
  and aerial users in cellular networks},'' in \emph{Proc. IEEE Global Commun.
  Conf. (Globecom) Wkshps.}, Dec. 2017.

\bibitem{Nguyen18}
H.~C. Nguyen \emph{et~al.}, ``{How to ensure reliable connectivity for aerial
  vehicles over cellular networks},'' \emph{IEEE Access}, vol.~6, pp.
  12\,304--12\,317, 2018.

\bibitem{Liang18}
L.~Liu, S.~Zhang, and R.~Zhang, ``{Multi-beam UAV communication in cellular
  uplink: cooperative interference cancellation and sum-rate maximization},''
  submitted to \emph{IEEE Trans. Wireless Commun.} [Online] Available:
  https://arxiv.org/abs/1808.00189.

\bibitem{Mei18}
W.~Mei, Q.~Wu, and R.~Zhang, ``{Cellular-connected UAV: Uplink association,
  power control and interference coordination},'' submitted to \emph{IEEE
  Trans. Wireless Commun.} [Online] Available:
  https://arxiv.org/abs/1807.08218.

\bibitem{NOMA}
Z.~Ding, Y.~Liu, J.~Choi, Q.~Sun, M.~Elkashlan, I.~C. Lin, and H.~V. Poor,
  ``{Application of non-orthogonal multiple access in LTE and 5G networks},''
  \emph{IEEE Commun. Mag.}, vol.~55, no.~2, pp. 185--191, Feb. 2017.

\bibitem{Ding2017}
Z.~Ding, X.~Lei, G.~K. Karagiannidis, R.~Schober, J.~Yuan, and V.~K. Bhargava,
  ``{A survey on non-orthogonal multiple access for 5G networks: Research
  challenges and future trends},'' \emph{IEEE J. Sel. Areas Commun.}, vol.~35,
  no.~10, pp. 2181--2195, Oct. 2017.

\bibitem{Ding2018}
L.~Dai, B.~Wang, Z.~Ding, Z.~Wang, S.~Chen, and L.~Hanzo, ``{A survey of
  non-orthogonal multiple access for 5G},'' \emph{IEEE Commun. Surveys Tuts.},
  vol.~20, no.~3, pp. 2294--2323, 3rd Quart. 2018.

\bibitem{Jafar08}
V.~Cadambe and S.~Jafar, ``{Interference alignment and the degrees of freedom
  of the K-user interference channel},'' \emph{IEEE Trans. Inf. Theory},
  vol.~54, no.~8, pp. 3425--3441, Aug. 2008.

\bibitem{Luo07}
E.~Karipidis, N.~D. Sidiropoulos, and Z.~Q. Luo, ``Far-field multicast
  beamforming for uniform linear antenna arrays,'' \emph{IEEE Tran. Signal
  Process.}, vol.~55, no.~10, pp. 4916--4927, Oct. 2007.

\end{thebibliography}

\end{document}